\documentstyle[epsfig,aps,prc]{revtex}

\begin{document}
\title{Evaluation of cluster expansions and correlated
one-body properties of nuclei}
\author{ Ch.C. Moustakidis, S.E. Massen,  C.P. Panos, 
M.E. Grypeos}
\address{
Department of  Theoretical Physics, Aristotle University of Thessaloniki
GR-54006 Thessaloniki, Greece }
\author{and A.N. Antonov}
\address{Institute of Nuclear Research and Nuclear Energy,
Bulgarian Academy of Sciences, Sofia 1784, Bulgaria}
%\date{}
\maketitle
\begin{abstract}
Three different cluster expansions for the evaluation of correlated
one-body properties of s-p and s-d shell nuclei are compared. Harmonic 
oscillator wave functions and Jastrow type correlations are used, while 
analytical expressions are obtained for the charge form factor, density 
distribution, and momentum distribution by truncating the expansions
and using a standard Jastrow correlation function $f$. The harmonic 
oscillator parameter $b$ and the correlation parameter $\beta$ have been 
determined by a least-squares fit to the experimental charge form factors
in each case. The information entropy of nuclei
in position-space ($S_r$) and momentum-space ($S_k$) according to the
three methods are also calculated. It is found that the larger the 
entropy sum $S=S_r+S_k$ (the information content of the system) the smaller 
the values of $\chi^2$.
This indicates that  $S$ is a criterion of the quality of a given
nuclear model, according to the maximum entropy principle. Only
two exceptions to this rule, out of many cases examined, were found.
Finally an analytic expression for the so-called "healing" or "wound"
integrals is derived with the function $f$ considered, for any state 
of the relative two-nucleon motion and their values in certain cases
are  computed and compared.
\end{abstract}

%\newpage
%%%%%%%%%%%%%%%%%%%%%%%%%%%%%%%%%%%%%%%%%%%%%%%%%%%%%%%%%%%%%%%%%%%%%%%%%%%
\section{Introduction}
The effect of short-range correlations (SRC) to the one-body properties
of nuclei is an old but challenging and appealing problem.
In general, the account of SRC is important for the description
of the mean values of some two-body operators,  such as the  ground
state energy of nuclei but it is also of interest to investigate
the SRC contribution to simpler nuclear quantities related to
one-body operators such as the form factor (FF), density distribution
(DD) and momentum distribution (MD). It has been shown that mean-field
theories can not describe correctly MD and DD simultaneously \cite{Jaminon}
and the main features of MD  depend little on the effective mean-field
considered \cite{Casas87}.
The reason is that MD is sensitive to short-range and tensor nucleon-nucleon
correlations which are not included in the mean-field theories.  
We note however that the choice of a single particle potential having a 
short range repulsion could play a role in improving somehow the values of 
MD \cite{Ypsilantis}. 

The experimental evidence obtained from inclusive and exclusive electron
scattering on nuclei established the existence of a high-momentum component
for momenta $k > 2 \ {\rm fm}^{-1}$ \cite{Day87,Ji90,Ciofi89,Antonov88}.
It is well
known, that the independent-particle model (IPM) fails to reproduce the high
momentum transfer data from electron scattering in nuclei. That is,
the IPM is inadequate to reproduce satisfactorily the diffraction minima of 
the charge FF for high values of momentum transfer. 
Therefore, although single-particle potentials of the type mentioned above,
that is with a short-range repulsion lead to certain improvement, theoretical 
approaches which take into account SRC due to the character of the
nucleon-nucleon forces at small distances, are necessary to be developed.

In this effort, two main problems appear.
The first one is the type of SRC which must be incorporated to
the mean-field nucleon wave function and the second one is the type of
cluster-expansion to be used which is connected with the number
of simultaneously correlated nucleons.

In the present work we consider central correlations of Jastrow
type \cite{Jastrow55} while
three different  cluster expansions are considered. The first two types of
expansions, named FIY (Factor, Iwamoto and Yamada) \cite{Iwamoto 57}
and FAHT (Factor, Aviles, Hartogh and Tolhoek) \cite{Aviles58} respectively,
were developed by Clark and co-workers \cite{Clark67,Feenberg69}
while the third one named LOA (Low Order Approximation)
was derived by Gaudin, Gillespie and Ripka \cite{Gaudin71,DalRi82}.

The FIY expansion, truncated at the two-body terms, was used for the
calculation of the charge FF and DD \cite{Massen99} and MD
\cite{Moustakidis00}  in $s$-$p$ and $s$-$d$ shell nuclei while the LOA,
truncated at the two-body terms and including a part of the three-body term, 
was used for the calculation of the above one-body quantities in the closed 
shell nuclei $^{4}$He, $^{16}$O and $^{40}$Ca \cite{Stoitsov93}
as well as of the bound-states overlap functions, separation
energies and spectroscopic factors in $^{16}$O and $^{40}$Ca
\cite{Stoitsov96}.
The FAHT expansion, truncated at the two-body terms was used for the
evaluation of the charge FF \cite{Guardiola72} and nuclear
ground state energy  of $^{4}$He and $^{16}$O \cite{Guardiola79}.
In the present paper the FAHT expansion is used in addition for the
evaluation of  the FF, DD and MD in s-p and s-d shell nuclei.

The present work is, in a way, a generalization of Ref. \cite{Ciofi69} 
where a comparison
of various cluster expansions for the calculation of the charge FF
of $^{4}$He was made. In this generalization, the above mentioned
three types of expansions are applied and  compared for the one-body
characteristics of $s$-$p$ and $s$-$d$ shell nuclei.

The comparison of the three truncated expansions can be made, as usually, by
comparing $\chi^2$ (in computing the FF) i.e. the smaller the $\chi^2$, the 
better the quality of the corresponding expansion.
In the present work we introduce also an information-theoretical
criterion in addition to $\chi^2$. Information-theoretical methods 
\cite{Bialy75,Gadre84,Gadre85,Gadre87,Ohya93,Nagy96,Majer96,Panos97,Lalazi98,%
Massen98,Panos00}
play an important role for the study of quantum-many body systems.
It has been found in Ref. \cite{Massen98} that interesting properties of the
information entropy $S$  hold for various systems. For instance, it was shown
that $S=a+b\ln N$ where $N$ is the number of particles in nuclei, atomic
clusters and atoms.
In a previous work \cite{Lalazi98} it was found that the larger the $S$, the 
better the quality of the nuclear model. Here we apply this idea to compare 
various cluster expansions. It turns out that this is the case i.e. the 
larger the $S$ the smaller the values of $\chi^2$, for various nuclei and 
expansions,  with only two exceptions.

The paper is organized as follows. In Sec. II the general expressions
of the one-body density matrix (OBDM) for the three types of expansions
are given. Numerical results are reported and discussed in Sec. III,
while the summary of the present work is given in Sec. IV.
Finally, some details of the FAHT expansion as well as for the calculation
of healing integrals are given in Appendix I and II, respectively.

\section{Correlated one-body properties}
\subsection{General definitions}

The key of the description of the one-body properties of nuclei is the OBDM
$\rho ({\bf r},{\bf r}')$, which for a system of A identical particles
is defined \cite{Dirac30,Lowdin55} in terms of the complete
wave function  $\Psi({\bf r}_1,{\bf r}_2,...,{\bf r}_A) $ by
\begin{equation}
\rho ({\bf r},{\bf r}')= \int \Psi^{*}({\bf r},{\bf r}_2, \dots ,{\bf r}_A)
\Psi ({\bf r}',{\bf r}_2, \dots , {\bf r}_A)   d {\bf r}_2 \ldots
 d {\bf r}_A,
\end{equation}
where the integration is carried out over the radius vectors and summation
over spin and isospin variables is implied.

In the case where the nuclear wave function
$\Psi({\bf r}_1,{\bf r}_2, \ldots ,{\bf r}_A)$
can be expressed as a single Slater determinant depending on the
single-particle wave functions we have
\begin{equation}
\rho_{SD}({\bf r},{\bf r}')=
\sum_{i=1}^{A}\phi_i^{*}({\bf r}) \phi_i({\bf r}') .
\end{equation}
The diagonal elements of the OBDM give the
DD, $\rho({\bf r},{\bf r})=\rho({\bf r})$
while the FF is the Fourier transform of it
\begin{equation}
F({\bf q})=\int \exp [i{\bf q}  {\bf r}]  \rho({\bf r})   d{\bf r} ,
\label{F-T}
\end{equation}
and the MD is given by a particular Fourier transform of the OBDM
\begin{equation}
n({\bf k})=\frac{1}{(2\pi)^3}\int \exp[i{\bf k}({\bf r}-{\bf r}')]
\rho({\bf r},{\bf r}')  d{\bf r}  d{\bf r}' .
\label{M-D}
\end{equation}

The second moment of the DD is the mean square radius of the nucleus
while the  second moment of the MD is related to the mean kinetic energy.

We also define the information entropy sum
\begin{equation}
S=S_r+S_k  ,
\label{ie-sum}
\end{equation}
where
\begin{equation}
S_r=-\int \rho({\bf r}) \ln\rho({\bf r}) d{\bf r}
\end{equation}
is the information entropy in position-space and
\begin{equation}
S_k=-\int n({\bf k}) \ln{n({\bf k})} d{\bf k}
\end{equation}
is the information entropy in momentum-space.

$S$ is a measure of quantum-mechanical uncertainty and represents the 
information content of a probability distribution, in our case of the 
nuclear density and momentum distributions. In the present work, we 
employ in calculating $S$ a normalization to the number of particles A
for $\rho({\bf r})$ and $n({\bf k})$ .

\subsection{The cluster expansions of the one-body density matrix}
%%%%%%%%%%%%%%%%%%%%%%%%%%%%%%%%%%%%%%%%%%%%%%%%%%%%%%%%%%%%%%%%%

The trial wave function $\Psi$, which describes a correlated nuclear system,
can be written as (e.g. \cite{Bruckner55})
\begin{equation}
\Psi={\cal F}\Phi ,
\end{equation}
where $\Phi$ is a model wave function which is adequate to describe
the uncorrelated A-particle nuclear system and
${\cal F}$ is the operator which introduces SRC. $\Phi$ is chosen to be
a Slater determinant wave function, constructed by  single-particle wave
functions. Several restrictions can be made on the model operator ${\cal F}$
\cite{Clark79,Brink67}.
In the present work $\cal{F}$ is taken to be of the Jastrow-type
\cite{Jastrow55}
\begin{equation}
{\cal F}=\prod_{i<j}^{A}f(r_{ij}) ,
\end{equation}
where $f(r_{ij})$ is the state-independent correlation function
of the form
\begin{equation}
f(r_{ij})=1-\exp[-\beta({\bf r}_i-{\bf r}_j)^2]  .
\label{fr-ij}
\end{equation}

\subsubsection{Factor cluster expansion of Iwamoto-Yamada}
%%%%%%%%%%%%%%%%%%%%%%%%%%%%%%%%%%%%%%%%%%%%%%%%%%%%%%%%%

In the factor cluster expansion of Iwamoto-Yamada (FIY) the OBDM
takes the form \cite{Moustakidis00}
\begin{equation}
\rho_{FIY}({\bf r},{\bf r'})=N [ \langle {\bf O}_{\bf rr'}\rangle_1
-O_2({\bf r},{\bf r'},{\rm g}_1)-O_2({\bf r},{\bf r'},{\rm g}_2)+
O_2({\bf r},{\bf r'},{\rm g}_3) ],
\end{equation}
where
$N$ is the normalization factor, and the terms
$\langle {\bf O}_{\bf rr'}\rangle_1$
and  $O_2({\bf r},{\bf r'},{\rm g}_l)$ ($l=1,2,3$) have the general forms
\begin{equation}
\langle {\bf O}_{\bf rr'}\rangle_1=\rho_{SD}({\bf r},{\bf r}')
= \frac{1}{\pi} \sum_{nl} \eta_{nl} (2l+1)
 \phi^{*}_{nl}(r) \phi_{nl}(r') P_l(\cos \omega_{rr'} ),
\label{O1-1}
\end{equation}
and
\begin{equation}
O_2({\bf r},{\bf r'},{\rm g}_l)=
\int {\rm g}_l({\bf r},{\bf r'},{\bf r}_2)
[\rho_{SD}({\bf r},{\bf r'})\rho_{SD}({\bf r}_2,{\bf r}_2)
-\rho_{SD}({\bf r},{\bf r}_2)\rho_{SD}({\bf r}_2,{\bf r'})]  d{\bf r}_2 ,
\label{O22-1}
\end{equation}
where
\begin{eqnarray}
&& {\rm g}_1({\bf r},{\bf r'},{\bf r}_2) = \exp [-\beta (r^2+r_2^2)]
\exp [2\beta {\bf r} {\bf r}_2], \quad
{\rm g}_2({\bf r},{\bf r'},{\bf r}_2)= {\rm g}_1({\bf r'},{\bf r},{\bf r}_2) ,
\nonumber\\
&& {\rm g}_3({\bf r},{\bf r'},{\bf r}_2) =
\exp [-\beta (r^2+{r'}^2)] \exp [-2\beta r_2^2]
\exp [2\beta ({\bf r}+{\bf r'}){\bf r}_2] .
\end{eqnarray}

The term $O_2({\bf r},{\bf r}',{\rm g}_l)$, performing the
spin-isospin summation and the angular integration, takes the general form
\begin{eqnarray}
O_2({\bf r},{\bf r}',  {\rm g}_l)& = & 4 \sum_{n_i l_i,n_j l_j}
\eta_{n_i l_i} \eta_{n_j l_j} (2 l_i +1) (2 l_j +1 )  \nonumber \\
&  &\times \left[ 4 A_{n_il_in_jl_j}^{n_il_i n_jl_j,0 }
({\bf r},{\bf r}', {\rm g}_l) - \sum_{k=0}^{l_i +l_j}
\langle l_i 0 l_j 0 \mid k 0 \rangle^2
 A_{n_il_in_jl_j}^{n_jl_j n_il_i,k}({\bf r},{\bf r}',{\rm g}_l) \right],
\label{O22-g-3}
\end{eqnarray}
where
\begin{eqnarray}
A_{n_1l_1n_2l_2}^{n_3l_3n_4l_4,k}({\bf r},{\bf r}', {\rm g}_1)& =&
\frac{1}{4\pi}\phi^{*}_{n_1l_1}(r) \
\phi_{n_3l_3}(r') \ \exp[-\beta r^2] \ P_{l_3}(\cos\omega_{rr'})
 \nonumber \\
& &\times \int_{0}^{\infty}\phi^{*}_{n_2l_2}(r_2)  \phi_{n_4l_4}(r_2)
\exp[-\beta r_{2}^2] \ i_k (2 \beta r r_2)
 r_{2}^{2}   d r_2  ,
\label{A-O22-1}
\end{eqnarray}
and the matrix element
$A_{n_1l_1n_2l_2}^{n_3l_3n_4l_4,k}({\bf r},{\bf r}', {\rm g}_2)$
can be found from (\ref{A-O22-1}) replacing
${\bf r}\leftrightarrow {\bf r}'$ and  $n_1l_1\leftrightarrow n_3l_3$
while the matrix element corresponding to the factor $ {\rm g}_3$ can be found
from (\ref{A-O22-1}) replacing the factors $\exp[-\beta r^2]$,
$P_{l_3}(\cos\omega_{rr'})$ and $i_k (2 \beta r r_2)$ by the factors
$\exp[-\beta (r^2+r'^2)]$, $\Omega_{l_1l_3}^{k}(\omega_{rr'})$ and
$i_k (2 \beta |{\bf r}+{\bf r}'| r_2)$ respectively \cite{Moustakidis00}.
In the expressions  of the matrix elements
$A_{n_1l_1n_2l_2}^{n_3l_3n_4l_4,k}({\bf r},{\bf r}', {\rm g}_l)$,
$i_k (z)$ is the modified
spherical Bessel function and the factor $\Omega_{l_1l_3}^{k}(\omega_{rr'})$
depends on the directions of ${\bf r}$ and ${\bf r}'$.

\subsubsection{Factor cluster expansion of Aviles, Hartogh and Tolhoek}
%%%%%%%%%%%%%%%

In the factor cluster expansion of Aviles, Hartogh and Tolhoek (FAHT),
truncated at the two-body terms, the OBDM takes the form (details of 
the calculations are given in Appendix I),
\begin{eqnarray}
\rho_{FAHT}({\bf r},{\bf r'})&=&\frac{1}{A}\langle {\bf O}_{\bf rr'}\rangle_1
\nonumber \\
&& + (A-1)\left[ \frac{(A-1)\langle {\bf O}_{\bf rr'}\rangle_1-
O_2({\bf r},{\bf r'},{\rm g}_1)-O_2({\bf r},{\bf r'},{\rm g}_2)+
O_2({\bf r},{\bf r'},{\rm g}_3)}{A(A-1) -
\int [O_2({\bf r},{\bf r},{\rm g}_1)+O_2({\bf r},{\bf r},{\rm g}_2)-
O_2({\bf r},{\bf r},{\rm g}_3)] d{\bf r} }
-\frac{1}{A}\langle {\bf O}_{\bf rr'}\rangle_1 \right] ,
\label{faht-1}
\end{eqnarray}
where $\langle {\bf O}_{\bf rr'}\rangle_1$ and
$O_2({\bf r},{\bf r'},{\rm g}_l)$
are given again by Eqs. (\ref{O1-1}) and (\ref{O22-1}) respectively.
The FAHT expansion has the advantage that the normalization
is preserved term by term.

\subsubsection{Low order approximation}
%%%%%%%%%%%%%%%%%%%%%%%%%%%%%%%%%%%%%%%

In the low order approximation (LOA) of Gaudin et al \cite{Gaudin71}
the Jastrow wave function $\Psi$
of the nucleus was expanded in terms of the functions
$\tilde{g}=f^2(r_{ij})-1$ and $h=f(r_{ij})-1$ and was truncated up to
the second order of $h$ and the first order of $\tilde{g}$. This expansion
contains one- and two-body terms and a part of the three-body term
which was chosen so that the normalization of the wave function was
preserved. In LOA the OBDM takes the form \cite{Gaudin71,DalRi82,Stoitsov93}
\begin{equation}
\rho_{LOA}({\bf r},{\bf r'})=\frac{1}{A}[\langle {\bf O}_{\bf rr'}\rangle_1
-O_2({\bf r},{\bf r'},{\rm g}_1)-O_2({\bf r},{\bf r'},{\rm g}_2)+
O_2({\bf r},{\bf r'},{\rm g}_3)
+2O_{3}({\bf r},{\bf r'},\beta)-O_{3}({\bf r},{\bf r'},2\beta)],
\end{equation}
where
$\langle {\bf O}_{rr'}\rangle_1$ and $O_2({\bf r},{\bf r'},{\rm g}_l)$
are given again by Eqs. (\ref{O1-1}) and (\ref{O22-1}) respectively and
the three-body term $O_{3}({\bf r},{\bf r'},z)$ ($z=\beta,\ 2\beta$)
has the form
\begin{equation}
O_{3}({\bf r},{\bf r'},z)=
\int {\rm g}({\bf r}_2,{\bf r}_3, z) \rho_{SD}({\bf r},{\bf r}_2)
[\rho_{SD}({\bf r}_2,{\bf r'})\rho_{SD}({\bf r}_3,{\bf r}_3)-
\rho_{SD}({\bf r}_2,{\bf r}_3)\rho_{SD}({\bf r}_3,{\bf r'})] \
 d{\bf r}_2  d{\bf r}_3 ,
\end{equation}
where
\begin{equation}
{\rm g}({\bf r}_2,{\bf r}_3, z)=
\exp[-z(r_{2}^2+r_{3}^2 -2{\bf r_2}{\bf r_3})].
\end{equation}

The term $O_3({\bf r},{\bf r}',z)$, performing the
spin-isospin summation and the angular integration, takes the general form
\small{
\begin{eqnarray}
O_3({\bf r},{\bf r}',  z)&=& 4 \sum_{n_i l_i,n_j l_j, n_k l_k}
\eta_{n_i l_i} \eta_{n_j l_j} \eta_{n_k l_k}
(2 l_i +1) \times \nonumber \\
&&\left[4  (2 l_k +1 ) \delta_{l_i l_j}
 A_{n_il_in_jl_j n_k l_k}^{n_jl_j n_il_i,n_kl_k,0 }
({\bf r},{\bf r}', z) -
(2 l_j +1 ) \delta_{l_i l_k} \sum_{k'=0}^{l_i +l_j}
\langle l_i 0 l_j 0 \mid k'0 \rangle^2
A_{n_il_in_jl_j n_kl_k}^{n_kl_k n_il_i n_jl_j,k'}({\bf r},{\bf r}',z) \right],
\label{O3-z}
\end{eqnarray}}
\normalsize
where
\begin{eqnarray}
A_{n_1l_1n_2l_2n_3l_3}^{n_4l_4n_5l_5,n_6l_6,k'}({\bf r},{\bf r}', z)& =&
\frac{1}{4\pi}\phi^{*}_{n_1l_1}(r) \phi_{n_4l_4}(r')
P_{l_1}(\cos\omega_{rr'})
\times \int_0^{\infty}\phi^{*}_{n_2l_2}(r_2) \phi_{n_5l_5}(r_2)
\exp[-z r_{2}^{2}] r_{2}^{2} d r_{2}
\nonumber\\
& &\times \int_{0}^{\infty}\phi^{*}_{n_3l_3}(r_3) \ \phi_{n_6l_6}(r_3)
\exp[-z r_{3}^2] \exp[2 z r_2 r_3]
\ r_{3}^{2} d r_3   \ .
\label{A-O3}
\end{eqnarray}

Expressions (\ref{O1-1}), (\ref{O22-g-3}) and (\ref{O3-z}) were derived for
the closed shell nuclei with $N=Z$ where $\eta_{nl}$ is 0 or 1. For the
open shell nuclei (with $N=Z$) we use the same expressions where now
$0\le \eta_{nl} \le 1$. 
The normalization is preserved for the closed shell nuclei in all the 
expansions.
In the case of the open shell nuclei the normalization is preserved 
(in the above formalism) for FIY and FAHT expansions. In the case of LOA,
in which the number of particles is also conserved \cite{Vanneck97}, 
particular attention has to be paid in each open shell nucleus.

It is noted that the general expressions of the two- and three-body terms
of the density matrix
given by Eqs. (\ref{O22-g-3}) and (\ref{O3-z}) are also valid for the 
expansions of the DD, FF and MD. The only difference is the expressions of 
the matrix elements $A$ which have to be used. 
For the DD they are found from (\ref{A-O22-1}) putting ${\bf r'} = {\bf r}$, 
while the ones of the FF follow from Eq. (\ref{F-T}) replacing 
$\rho({\bf r})$ by $A({\bf r},{\bf r})$ and for the MD they follow from 
Eq. (\ref{M-D}) replacing $\rho({\bf r}, {\bf r'})$ by $A({\bf r},{\bf r'})$.

In the case when the model wave function $\Phi$ is constructed from
harmonic oscillator (HO) wave
functions, analytical expressions of the various terms of the DD, FF and
MD for any $N=Z$ $s$-$p$ and $s$-$d$ shell nuclei can be found for FIY and
FAHT while in the case of LOA analytical expressions of the closed
shell-nuclei in the same region can be found. These expressions which depend on
the
HO parameter $b$ and the correlation parameter $\beta$ are given in Refs.
\cite{Massen99,Moustakidis00,Stoitsov93} for FIY and LOA while the ones
for FAHT can be found easily from the other expansions.

\section{Results and Discussion}
%%%%%%%%%%%%%%%%%%%%%%%%%%%%%%%

The three expansions, mentioned in Sec. II, have been used for the analytical
calculations of the DD, MD and charge FF as well as for the calculation of
the information entropy sum defined by Eq. (\ref{ie-sum}). The HO parameter
$b$ and the SRC parameter $\beta$ in the three cases have been determined,
for each nucleus separately, by a least squares fit to the experimental
charge FF as in Ref. \cite{Massen99} (using the same expression for 
$\chi^2$).
The center-of-mass correction has been taken into account by a Tassie-Barker 
factor \cite{Tassie58} while those for the finite proton size and the 
Darwin-Foldy relativistic correction through the Chandra and Sauer 
approximation \cite{Chandra76}.
They are not taken into account in the calculations of DD and MD 
to obtain the information entropy sum (and in the plots of MD).

The variation with A of the  best fit values of the parameters $b$ and 
$\beta$ for each of the three expansions is shown in Fig. 1 where $b$ and 
$\beta$ versus the mass number A have been plotted for various $s$-$p$ and 
$s$-$d$ shell nuclei.
It is seen that these parameters have the same behaviour in FIY and FAHT
expansions. In the case of LOA expansion, which has been used only for
$^4$He, $^{16}$O and $^{40}$Ca the variation of the parameters
seems to be the same. From Fig. 1b it is seen also that the SRC parameter
$\beta$ has larger values in the open shell nuclei ($^{12}$C, $^{24}$Mg,
$^{28}$Si and $^{32}$S) than in the closed shell ones, indicating that
there should be a shell effect in the case of closed
shell nuclei.

In this work we compare different expansions on the example of MD for closed
and open shell nuclei. The reason for this is that the high-momentum
component of $n(k)$ is very sensitive to the extent to which nucleon
correlations are accounted for in a given correlation method and in various
approximations. The effect of different expansions on the form factors can
be seen comparing the values of $\chi^2$ for the various expansions
and nuclei.

The MD for the closed shell nuclei $^4$He, $^{16}$O and $^{40}$Ca,
calculated with the best fit values of the parameters and for the
three expansions, are shown in Fig. 2. It is seen that the inclusion of SRC
increases considerably the high momentum component of $n(k)$.
It has the same slope up to $2\ {\rm fm}^{-1}$ for the three expansions.
In the region $2 \ {\rm fm}^{-1} < k < 5 \ {\rm fm}^{-1}$ the slope seems to
be a little different.
FIY gives a larger contribution in the high momentum component than FAHT and
LOA which give the same contribution in this region.
The same behaviour of $n(k)$ has been observed in the open
shell nuclei as can be seen from Fig. 3. Here we would like to note that
in general, a more realistic description of MD requires the inclusion of
tensor correlations in the theoretical scheme.

In the previous analysis, the nuclei $^{24}$Mg, $^{28}$Si and $^{32}$S
were treated as $1d$ shell nuclei, that is, the occupation probability
of the $2s$ state was taken to be zero. The formalism of the
expansions FIY and FAHT
has the advantage that the occupation probabilities
of the various states can be treated as free parameters in the
fitting procedure of the charge FF. Thus, the analysis can be made with
more free parameters.
For that reason we considered, as in Ref. \cite{Moustakidis00}
the cases FIY$^*$ and FAHT$^*$ in which the occupation
probability $\eta_{2s}$ of the nuclei $^{24}$Mg, $^{28}$Si and $^{32}$S
was taken to be a free parameter together with the parameters $b$ and
$\beta$. We found that in both expansions the $\chi^2$ values become smaller,
compared to those of cases FIY and FAHT  and the $A$ dependence of the
parameter $\beta$, as can be seen from Fig. 1b, is not so strong as before.
Also the values of $\eta_{2s}$ found in the fit and the values of
$\eta_{1d}$ found through the relation
$\eta_{1d} = [(Z-8) - 2 \eta_{2s}]/10$,
are very close for both expansions in each nucleus.

Our best fit values of the parameters and the values of $\chi^2$ for the
various nuclei under consideration and for the three expansions
as well as for the HO case (that is when SRC are not included) are shown in
Table I. From the values of $\chi^2$ we conclude that the three
expansions give similar values of $\chi^2$. The FIY and FAHT expansions
have almost the same $\chi^2$ values.
They differ less than  $ 2 \%$ in the two expansions in each nucleus.
In most cases the $\chi^2$ values corresponding to FIY (or FIY$^*$) are
smaller. There are two cases ($^{12}$C and $^{28}$Si) when
the FAHT or FAHT$^*$ expansion gives smaller $\chi^2$ value and one
case ($^{16}$O) when LOA gives smaller $\chi^2$ value.

In addition, we verify the information-theoretic
criterion for comparing the quality of the three expansions. It is seen
in Table I that almost in all cases, the larger the $S$ the smaller the
$\chi^2$. Both methods of comparison ($S$ and $\chi^2$) show that the FIY
(or FIY$^*$) expansion is better than the FAHT and LOA for
$^{4}$He, $^{24}$Mg, $^{32}$S and $^{40}$Ca.
For $^{16}$O the LOA is the best. There are only two exceptions to this rule 
i.e. in $^{12}C$ for cases FIY and FAHT and in $^{28}$Si for cases 
FIY$^*$ and FAHT$^*$.
In $^{12}C$ $\chi^2$ is smaller in FAHT and we expect $S$ to
be larger than in FIY while in $^{28}$Si $\chi^2$ is smaller in FIY$^*$
and we expect $S$ to be larger than in FAHT$^*$. These are two
exceptions to our rule. It should be noted also that in these two exceptions
the difference in the $\chi^2$ values for the two expansions
in both nuclei is less than $1\%$.

Finally, we consider the so-called "healing" or "wound" integrals,
denoted here as $w^{2}_{nl}$ \cite{Brink67,Massen96} for the various 
states of the relative two-nucleon motion, pertinent
to the closed shell nuclei of Table I and in each case, that is
in each of the cluster expansions FIY, FAHT and LOA. The values of these
integrals express in a way the "amount of correlations" introduced to
each state of the relative two-nucleon motion.  The healing integrals
(for a state independent correlation function $f(r)$, such as the one
given by (\ref{fr-ij})) are defined as follows
\begin{equation}
w_{nl}^{2}=\int_{0}^{\infty}|\psi_{nl}(r)-\phi_{nl}(r)|^2 dr ,
\label{heal-int}
\end{equation}
where $\phi_{nl}(r)$ is the (normalized to unity), uncorrelated (HO) radial
relative wave function and $\psi_{nl}(r)$ the corresponding, normalized
to unity, correlated one: $\psi_{nl}(r)=N_{nl}f(r)\phi_{nl}(r)$,
where $N_{nl}$, the normalization factor of $\psi_{nl}(r)$, is given by
\begin{equation}
N_{nl}=\left[\int_{0}^{\infty}f^2(r)\phi_{nl}^2(r) dr \right]^{-1/2} .
\label{norm-healing}
\end{equation}

It is interesting to note that with the correlation function (\ref{fr-ij})
the healing integrals can be calculated analytically for every state
$nl$. Some details are given in Appendix II. As one expects, these integrals 
depend on both, the HO parameter b and the correlation parameter $\beta$.
We may note, however, that their dependence on them is
only through the dimensionless product
$y=2 \beta b^2$ (see expression (\ref{Inl-Jaco}) of Appendix II).
 
In Table II the values of the parameters $b$, $\beta$ and 
$\tilde{y}=\beta b^2$
for each closed shell nucleus and cluster expansion considered, are 
displayed along with the corresponding values of $w_{nl}^2$ for certain 
relative state in the $s$-$p$ and $s$-$d$ closed shell nuclei. 
It is seen from the results in this table that the values of $w^{2}_{nl}$, 
for each of the relative states $(nl)$ involved in each nucleus, are 
smaller when $w^{2}_{nl}$ is obtained with the FIY expansion and larger 
when obtained with the LOA. 
Furthermore, for each nucleus and expansion the values of $w^{2}_{nl}$ of 
the nodless (n=0) states decrease as the value of $l$ increases, the 
correlations having less effect to these higher $l$-states, because of 
the existing centrifugal (repulsive) term of the HO potential.
The values of $w_{n0}^2$ increase when $n=1$ or $n=2$ in comparison 
with those of $w_{00}^2$.

\section{Summary}
In the present work, a systematic study of the effect of SRC
on one-body properties of  $sp$ and $sd$ shell nuclei
has been made evaluating three different cluster expansions.
The HO parameter $b$ and the SRC parameter $\beta$
have been determined by a least-squares fit to the experimental charge FF.

The comparison of the three expansions on the example of the MD and the FF
shows that they can be considered as equivalent expansions. It is found that,
when the calculations are made with the best fit values of the parameters,
these expansions reproduce the diffraction minima of the FF in the correct
place and they give similar MD for all the nuclei we have considered.
The inclusion of SRC increases considerably the high momentum component of
$n(k)$.

The FIY and FAHT expansions have been used both for closed and for open 
shell nuclei while the occupation probabilities can be treated as free 
parameters together with the parameters $b$ and $\beta$ in the fitting 
procedure of the FF. In LOA such calculations are in progress.

In addition, the information entropy sum has been calculated
according to the three methods compared in the present work.
It was found almost in all of the numerous cases (different expansions 
and nuclei), that 
the larger the $S$, the smaller the $\chi^2$. That is $S$ could be
used as a criterion for the quality of a given nuclear model.
We found only two exceptions to this rule. In these two exceptions the
difference of the $\chi^2$ values is less then $1\%$.

Finally, attention was paid to the "healing" or "wound" integrals $w_{nl}^2$
of the relative two nucleon states. A convenient analytic expression of
$w_{nl}^2$ with correlation function (\ref{fr-ij}) was derived for any 
relative state $nl$. Their values were computed in a number of states
with that expression and were also discussed.

%%%%%%%%%%%%%

\section{Appendix I}
%%%%%%%%%%%%%%%%%%%%

In this appendix, we give some details about the FAHT expansion.
We define the correlated wave function as
\begin{equation}
\Psi=\prod_{i<j}^{A} f({\bf r}_i,{\bf r}_j) \Phi ,
\end{equation}
where $f({\bf r}_i,{\bf r}_j)$ is the Jastrow correlation function and
$\Phi$ is a Slater determinant wave function.
To built up the cluster expansion, we start,
following Ref. \cite{Guardiola79},
from the A-body integrals $J_A(\lambda)$ defined as
\begin{equation}
J_{A}(\lambda)=\frac{1}{A(A-1) \cdots 1} \sum_{i_1 \ldots i_A}^{A}
\langle \phi_{i_1} \ldots  \phi_{i_A}|
\prod_{i<j}^{A}f({\bf r}_i,{\bf r}_j) {\bf O}_1(A) e^{\lambda {\bf O}_2(A)}
\prod_{i<j}^{A} f({\bf r}_i',{\bf r}_j')|
\phi_{i_1}' \ldots  \phi_{i_A}' \rangle_a \ ,
\end{equation}
where the sum over the states $i_1,i_2, \ldots , i_A$
has no restrictions and extends over all one-particle states and
$\alpha$ stands for the antisymmetrization.
The operators ${\bf O}_1(A)$ and ${\bf O}_2(A)$ have the forms
\begin{eqnarray}
{\bf O}_1(A)&=&\prod_{i=1}^{A} \delta({\bf r}_i-{\bf r}_i'), \nonumber\\
{\bf O}_2(A)&=& \frac{1}{\prod_{i=i}^{A} \delta({\bf r}_i-{\bf r}_i')}
\sum_{i=1}^{A} \ \delta({\bf r}_i-{\bf r})
\delta({\bf r}_i' -{\bf r}')\prod_{j \neq i}^{A}\delta({\bf r}_j-{\bf r}_j') .
\end{eqnarray}
The OBDM $\rho_{FAHT}({\bf r},{\bf r'})$, normalized to $A$, is defined as
\begin{equation}
\rho_{FAHT}({\bf r},{\bf r'})=
\left[ \frac{{\rm d}\ln J_A(\lambda)}{{\rm d} \lambda} \right]_{\lambda=0} .
\label{faht-ap}
\end{equation}
We introduce the n-body integrals $J_n(\lambda)$ defined as
\begin{equation}
J_{n}(\lambda)=\frac{1}{A(A-1) \cdots  (A-n+1)}
\sum_{i_1 \ldots i_n}^{n}
\langle \phi_{i_1} \ldots  \phi_{i_n}|
\prod_{i<j}^{n}f({\bf r}_i,{\bf r}_j) {\bf O}_1(n) e^{\lambda {\bf O}_2(n)}
\prod_{i<j}^{n} f({\bf r}_i',{\bf r}_j')|
\phi_{i_1}' \ldots  \phi_{i_n}' \rangle_a \ .
\end{equation}
The cluster integrals $\Im_{n}\ (n=1,2,\ldots A)$ are defined through the
successive application of the equation
\begin{equation}
J_{n}=\prod_{k=1}^{n}
\Im_{k}^{{\tiny \left(  \begin{array}{c}
n \\ k \end{array}  \right)\normalsize}}=
\Im_{1}^{{\tiny \left(  \begin{array}{c}
n \\ 1 \end{array}  \right)\normalsize}} \
\Im_{2}^{{\tiny\left(  \begin{array}{c}
n \\ 2 \end{array}  \right)\normalsize}} \
\cdots
\Im_{n}^{{\tiny \left(  \begin{array}{c}
n \\ n \end{array}  \right)\normalsize}}, \quad n=1,2,\ldots , A .
\label{cluster}
\end{equation}
For example, for $n=1$ and $n=2$ it gives
\begin{equation}
\Im_{1}=J_1\nonumber, \quad
\Im_{2}=\frac{J_2}{J_1^2} \ .
\end{equation}
The last of Eqs. (\ref{cluster}), which corresponds to
$n=A$ is the quantity we are interested in
\begin{equation}
J_A=\prod_{n=1}^{A}
\Im_{n}^{{\tiny \left(  \begin{array}{c}
A \\ n \end{array}  \right)\normalsize}}
\equiv
\Im_{1}^{{\tiny \left(  \begin{array}{c}
A \\ 1 \end{array}\right)\normalsize}}
\Im_{2}^{{\tiny \left(  \begin{array}{c}
A \\ 2 \end{array}  \right)\normalsize}}
\Im_{3}^{{\tiny \left(  \begin{array}{c}
A \\ 3 \end{array}  \right)\normalsize}} \cdots
\Im_A \ .
\end{equation}
If the factor-cluster expansion is limited to the two-body term (assuming
that the remaining cluster integrals are equal to unity \cite{Guardiola79}), 
then
\begin{equation}
J_A \approx \Im_{1}^{{\tiny \left(  \begin{array}{c}
A \\ 1 \end{array}\right)\normalsize}}
\Im_{2}^{{\tiny \left(  \begin{array}{c}
A \\ 2 \end{array}  \right)\normalsize}}.
\label{J-approx}
\end{equation}

From Eqs. (\ref{faht-ap}) and (\ref{J-approx}) we have
\begin{equation}
\rho_{FAHT}({\bf r},{\bf r'}) =
\left(  \begin{array}{c}
   A \\ 1 \end{array}\right)
\left[\frac{1}{J_1}\frac{{\rm d}J_1}{{\rm d} \lambda} \right]_{\lambda=0}+
\left(  \begin{array}{c}
   A \\ 2 \end{array}  \right)
\left[ \frac{1}{J_2}\frac{{\rm d} J_2}{{\rm d} \lambda} -
2 \frac{1}{J_1}\frac{{\rm d} J_1}{{\rm d} \lambda} \right]_{\lambda=0} ,
\end{equation}
where
\begin{equation}
J_1(\lambda)=\frac{1}{A}\sum_{i_1=1}^{A}
\langle \phi_{i_1}({\bf r}_1) | {\bf O}_1(1) e^{\lambda {\bf O}_2(1)}|
\phi_{i_1}({\bf r}_1')\rangle ,
\end{equation}
and
\begin{equation}
J_2(\lambda)=\frac{1}{A(A-1)} \sum_{i_1,i_2}^{A}
\langle \phi_{i_1}({\bf r}_1) \phi_{i_2}({\bf r}_2) |
f({\bf r}_1,{\bf r}_2) {\bf O}_1(2) e^{\lambda {\bf O}_2(2)}
f({\bf r}_1',{\bf r}_2')|
\phi_{i_1}({\bf r}_1') \phi_{i_2}({\bf r}_2')\rangle_a \ .
\end{equation}

After some algebra we obtain
\begin{eqnarray}
&&J_1(0)=1 ,  \nonumber\\
&&J_2(0)=\frac{1}{A(A-1)}\left[\frac{}{} A(A-1)-\int
[O_2({\bf r},{\bf r},g_1)+O_2({\bf r},{\bf r},g_2)-
O_2({\bf r},{\bf r},g_3)]{\rm d}{\bf r}\right] ,  \nonumber\\
&&\left[\frac{{\rm d} J_1}{{\rm d} \lambda}\right]_{\lambda=0}=
\frac{1}{A}\langle {\bf O}_{rr'}\rangle_1   , \nonumber\\
&&\left[\frac{{\rm d} J_2}{{\rm d} \lambda}\right]_{\lambda=0}=
\frac{2}{A(A-1)}\left[\frac{}{}(A-1)\langle {\bf O}_{rr'}\rangle_1-
O_2({\bf r},{\bf r'},g_1)-O_2({\bf r},{\bf r'},g_2)+
O_2({\bf r},{\bf r'},g_3)\right] ,
\end{eqnarray}
where the terms $\langle {\bf O}_{rr'}\rangle_1$ and
$O_2({\bf r},{\bf r'},{\rm g}_l)$ have been defined in Sec. II.

Finally, the $\rho_{FAHT}({\bf r},{\bf r'})$, normalized to unity,
becomes
\begin{eqnarray}
\rho_{FAHT}({\bf r},{\bf r'})&=&\frac{1}{A}\langle {\bf O}_{rr'}\rangle_1
\\
& & + (A-1)\left[\frac{}{}\frac{(A-1)\langle {\bf O}_{rr'}\rangle_1-
O_2({\bf r},{\bf r'},g_1)-O_2({\bf r},{\bf r'},g_2)+
O_2({\bf r},{\bf r'},g_3)}{A(A-1)-\int
[O_2({\bf r},{\bf r},g_1)+O_2({\bf r},{\bf r},g_2)-
O_2({\bf r},{\bf r},g_3)] d{\bf r}}
-\frac{1}{A}\langle {\bf O}_{rr'}\rangle_1\right] \nonumber .
\end{eqnarray}

%%%%%%%%%%%%%%%%%%%%%%%%
\section{Appendix II}
%%%%%%%%%%%%%%%%%%%%%

The healing integral defined by(\ref{heal-int}) is written as follows
\cite{Massen96}
\begin{equation}
w_{nl}^2 = 2 \left[ 1 + N_{nl} \left( I_{nl}(b,\beta ) -1 \right) \right] ,
\label{heal-int2}
\end{equation}
where
\begin{equation}
I_{nl}(b,\beta) = \int_0^{\infty} \exp[-\beta r^2] \phi_{nl}^2(r) d r ,
\label{Inl-1}
\end{equation}
and the normalization factor $N_{nl}$ is given by (\ref{norm-healing}). This
factor can be easily expressed in terms of the integrals $I_{nl}(b,\beta)$ 
and $I_{nl}(b,2 \beta)$ by means of expression (\ref{fr-ij}) 
\begin{equation}
N_{nl} = \left[ 1 - 2 I_{nl}(b,\beta) + I_{nl}(b,2 \beta) \right] ^{-1/2} .
\label{Nnl-2}
\end{equation}

Thus, the analytical calculation of any healing integral $w_{nl}^2$ is
reduced to the calculation of two integrals of type (\ref{Inl-1}). 
The expression of $I_{nl}(b,2 \beta)$ follows immediately from the 
expression of $I_{nl}(b, \beta)$.

We use the general expression of the radial HO wave function 
(normalized to one as $\int_0^{\infty} \phi_{nl}^2 d r =1$)
in the form
\begin{equation}
\phi_{nl}(r)=\left( \frac{2 n!}{\Gamma(n+l+\frac{3}{2}) b_r}
\right)^{\frac{1}{2}} \left(\frac{r}{b_r}\right)^{l+1}
L_n^{l+\frac{1}{2}} \left( \frac{r^2}{b_r^2} \right) 
\exp\left[ \frac{-r^2}{2 b_r^2} \right] , 
\label{HO-wf}
\end{equation}
where $b_r$ is the HO parameter of the relative motion, which is related 
to the usual HO parameter $b$ by $b_r=\sqrt{2}b$ 
($b=(\hbar /m\omega)^{1/2}$).

Substituting expression (\ref{HO-wf}) into (\ref{Inl-1})
and using the transformation $r^2/b_r^2 = \xi$, $I_{nl}$ is written
\begin{equation}
I_{nl}(b,\beta)=\frac{n!}{\Gamma[n+l+3/2]} 
\int_0^{\infty} {\rm e}^{-(1+y)\xi} \xi^{l+1/2} 
\left[ L_n^{l+\frac{1}{2}} (\xi) \right]^2 d \xi,
\end{equation}
where $y=\beta b_r^2=2\beta b^2$.

Using formula 13 of \S 7.414 of Ref. \cite{Grads} $I_{nl}$ takes the form
\begin{equation}
I_{nl}(b,\beta)=(y-1)^n(y+1)^{-n-l-3/2} P_n^{(l+\frac{1}{2}, 0)}
\left(\frac{y^2+1}{y^2-1}\right),
\label{Inl-Jaco}
\end{equation}
where $P_n^{(a_1,a_2)}(z)$ the Jacobi polymomials. 
These may be easily expressed in terms of the Hypergeometric function
(see e.g. \S 8.962 of Ref. \cite{Grads}).
In the case of the nodless states (because $P_0^{(a_1,a_2)}(z)=1$) 
$I_{nl}$ takes the simple form
\begin{equation}
I_{0l}(b,\beta)=(y+1)^{-l-3/2},
\label{Inl-Jaco-0}
\end{equation}

By substituting $\beta \rightarrow 2\beta$, the expression of 
$I_{nl}(b,2\beta)$ follows immediately and therefore the analytic expression
of the $w_{nl}^2$ by means of the formulae (\ref{heal-int2}) and 
(\ref{Inl-Jaco}).
It is thus clear that the healing integral $w_{nl}^2$ for any state 
depends on correlation parameter $\beta$ and the HO one, only
through the product $y=2\beta b^2$. The expressions of $w_{nl}^2$ for the 
lower $n-$states follow also very easily.

%%%%%%%%%%%%%%

%\newpage

%\newpage
\vspace*{1.cm}

\begin{figure}
\label{b-fig}
\begin{center}
\begin{tabular}{lr}
{\epsfig{figure=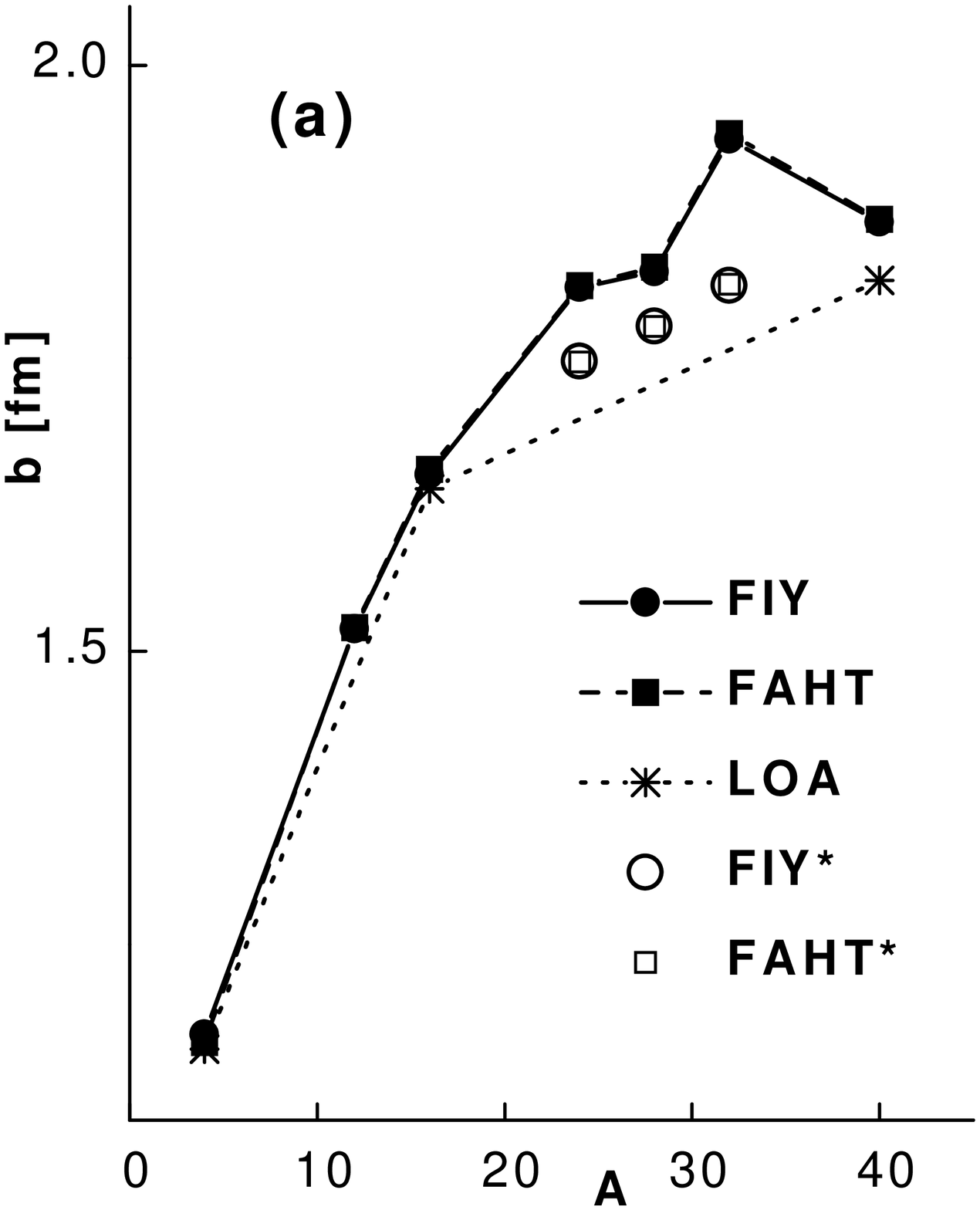,width=5cm} }&
{\epsfig{figure=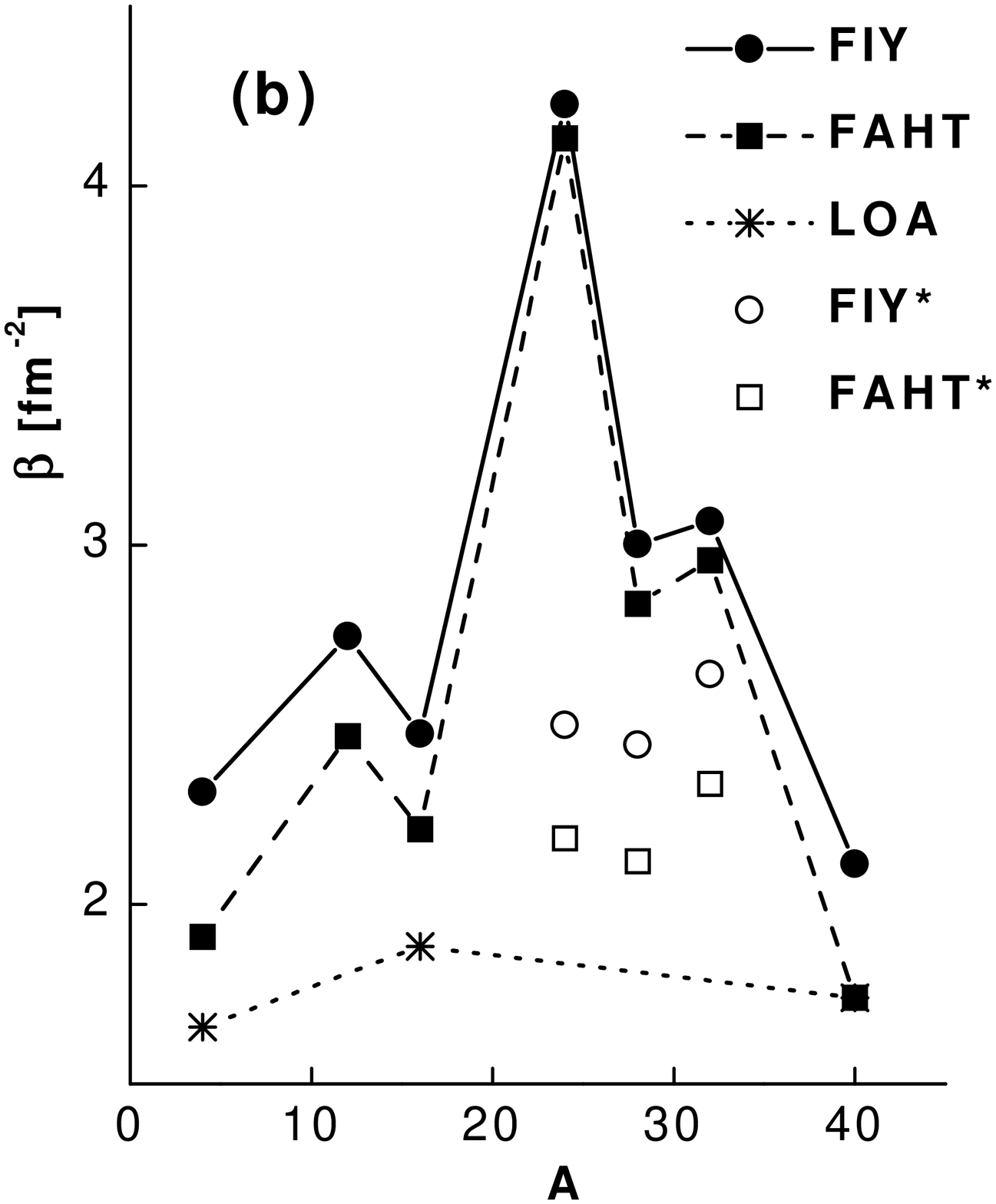,width=5cm} }\\
\end{tabular}
\end{center}
\caption{The harmonic oscillator parameter $b$ (a) and
correlation parameter $\beta$ (b) versus the mass number A
for the expansions FIY, FAHT and LOA. Cases FIY$^*$ and FAHT$^*$
(open circles and squares respectively) correspond
to the case when the occupation probability $\eta_{2s}$ is treated as a
free parameter.}
\end{figure}

%\newpage

\begin{figure}
\label{closedshell-fig}
\begin{center}
\begin{tabular}{ccc}
{\epsfig{figure=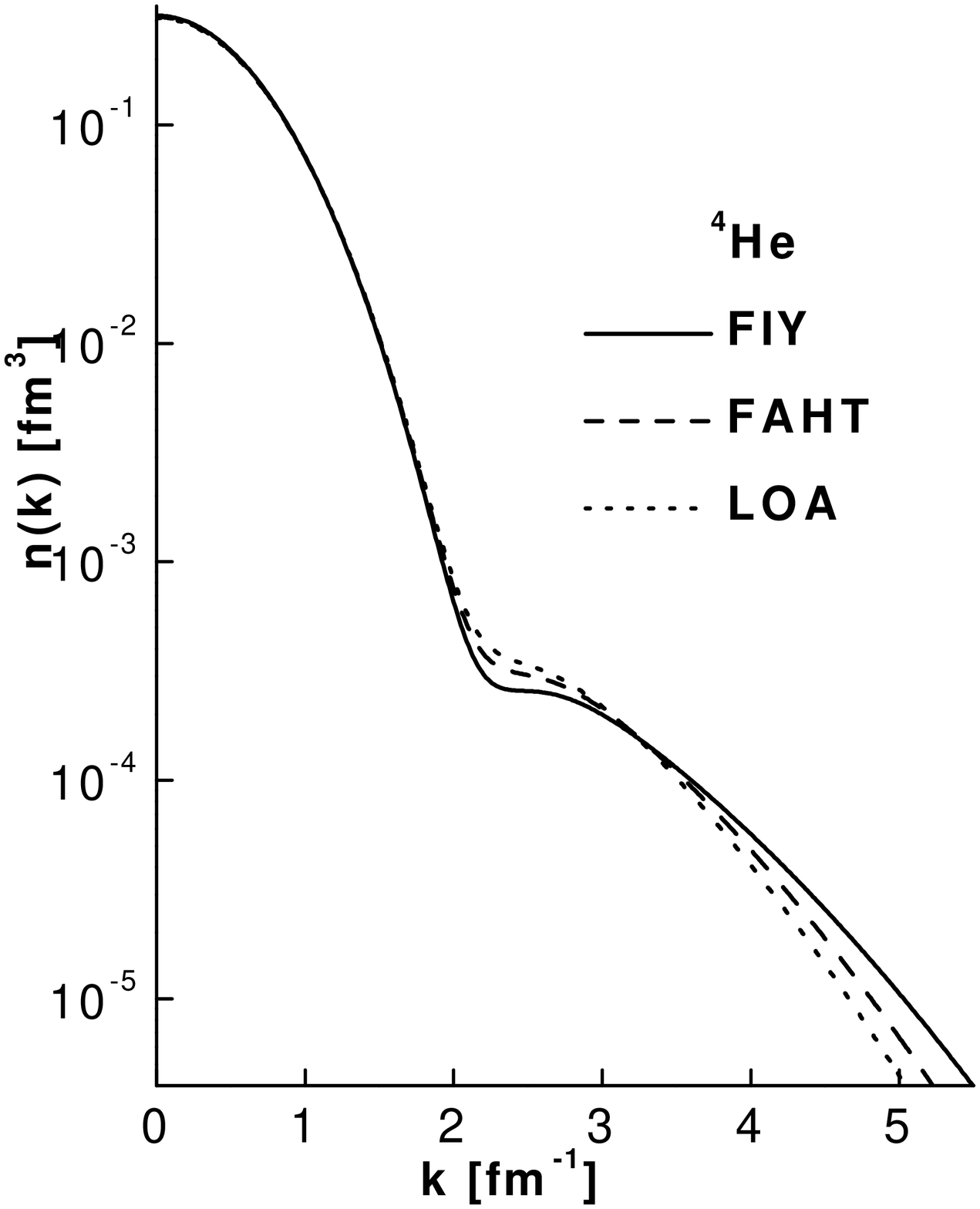,width=5.cm} }&
{\epsfig{figure=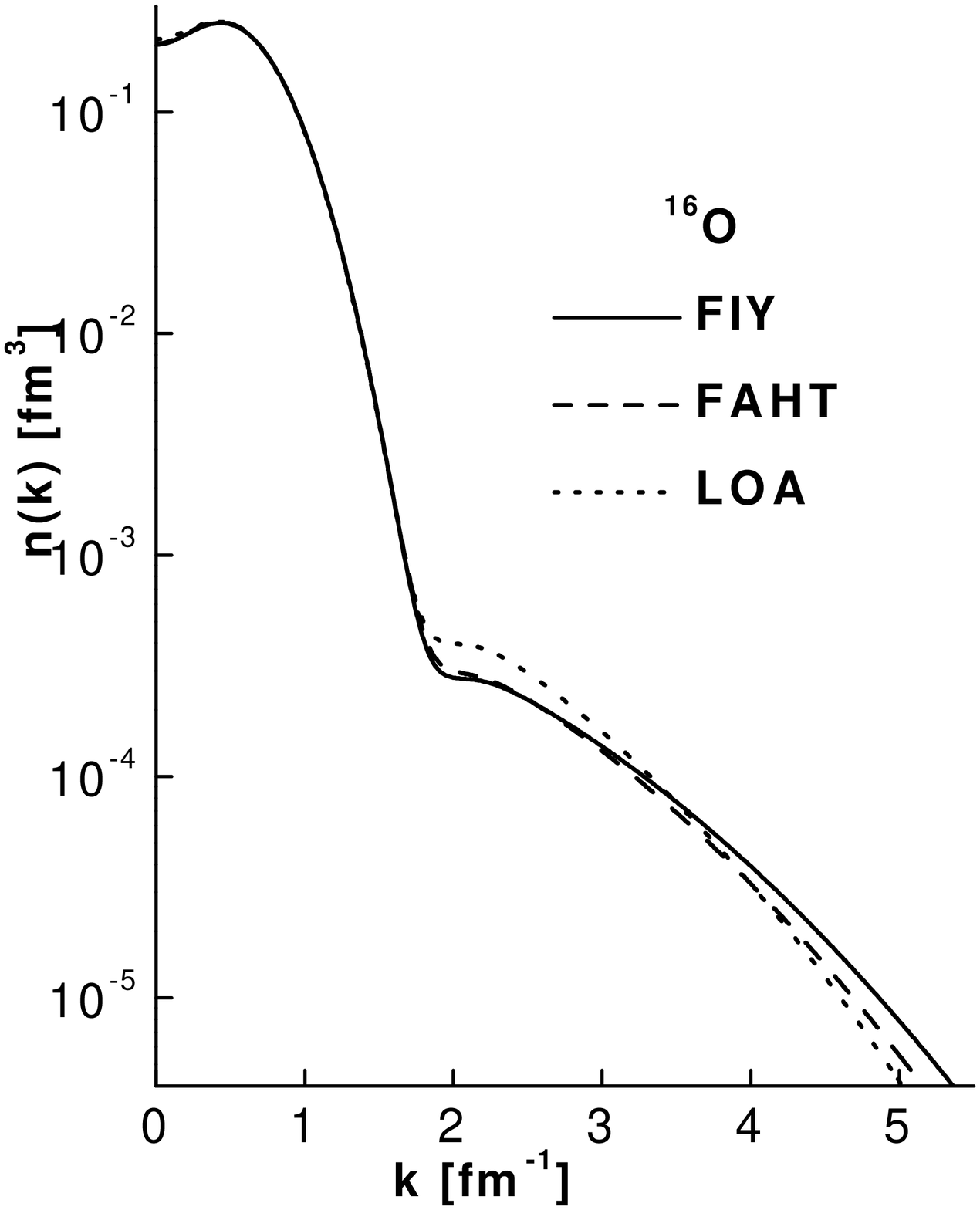,width=5.cm} }&
{\epsfig{figure=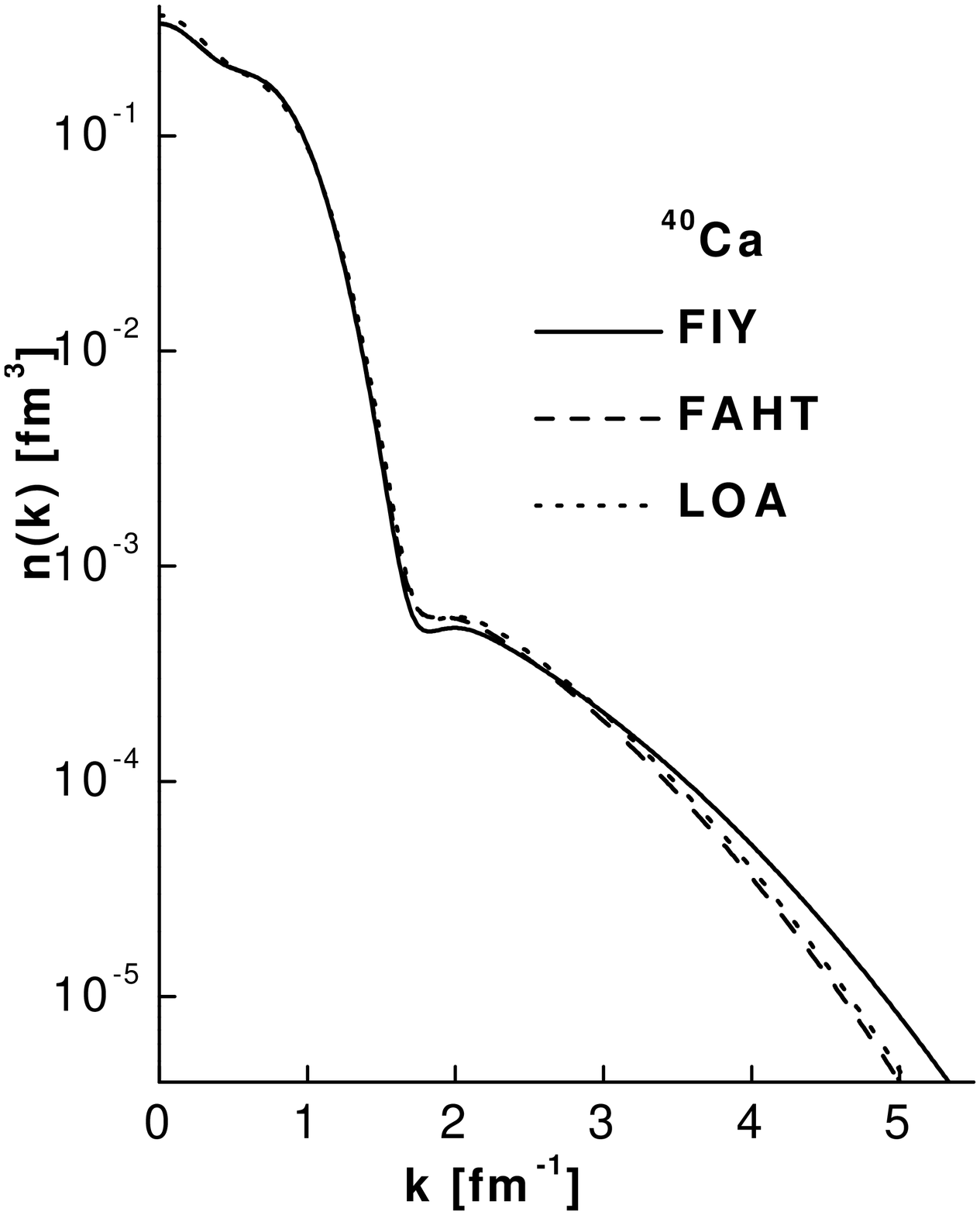,width=5.cm}}
\end{tabular}
\end{center}
\caption{The momentum distribution of the closed shell nuclei in the
three expansions, FIY (solid line), FAHT (dash line) and LOA (dot line).
The normalization is $\int n({\bf k}) d {\bf k} =1$.}
\end{figure}

%\newpage

\begin{figure}
\label{openshell-fig}
\begin{center}
\begin{tabular}{cc}
{\epsfig{figure=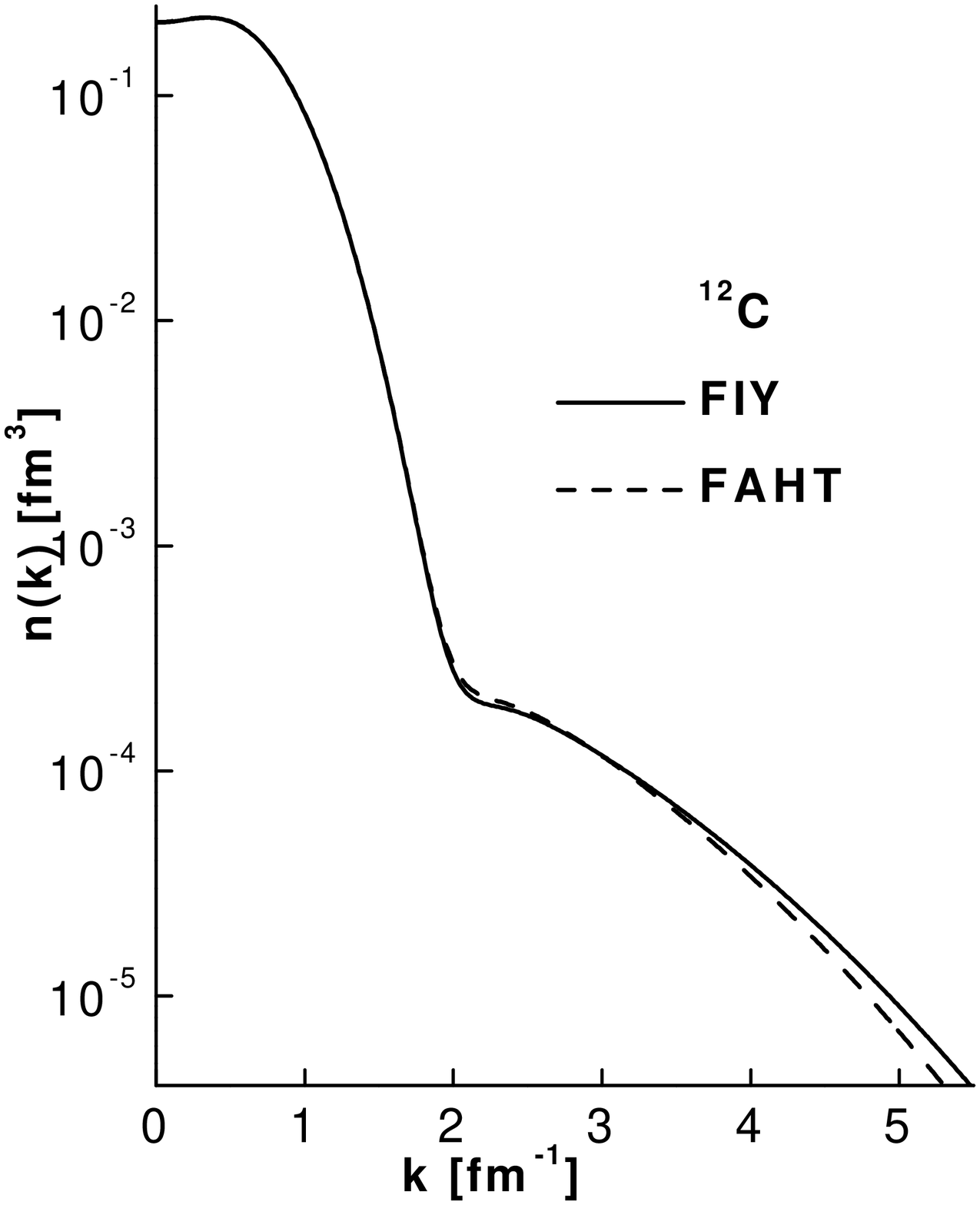,width=5cm} } &
{\epsfig{figure=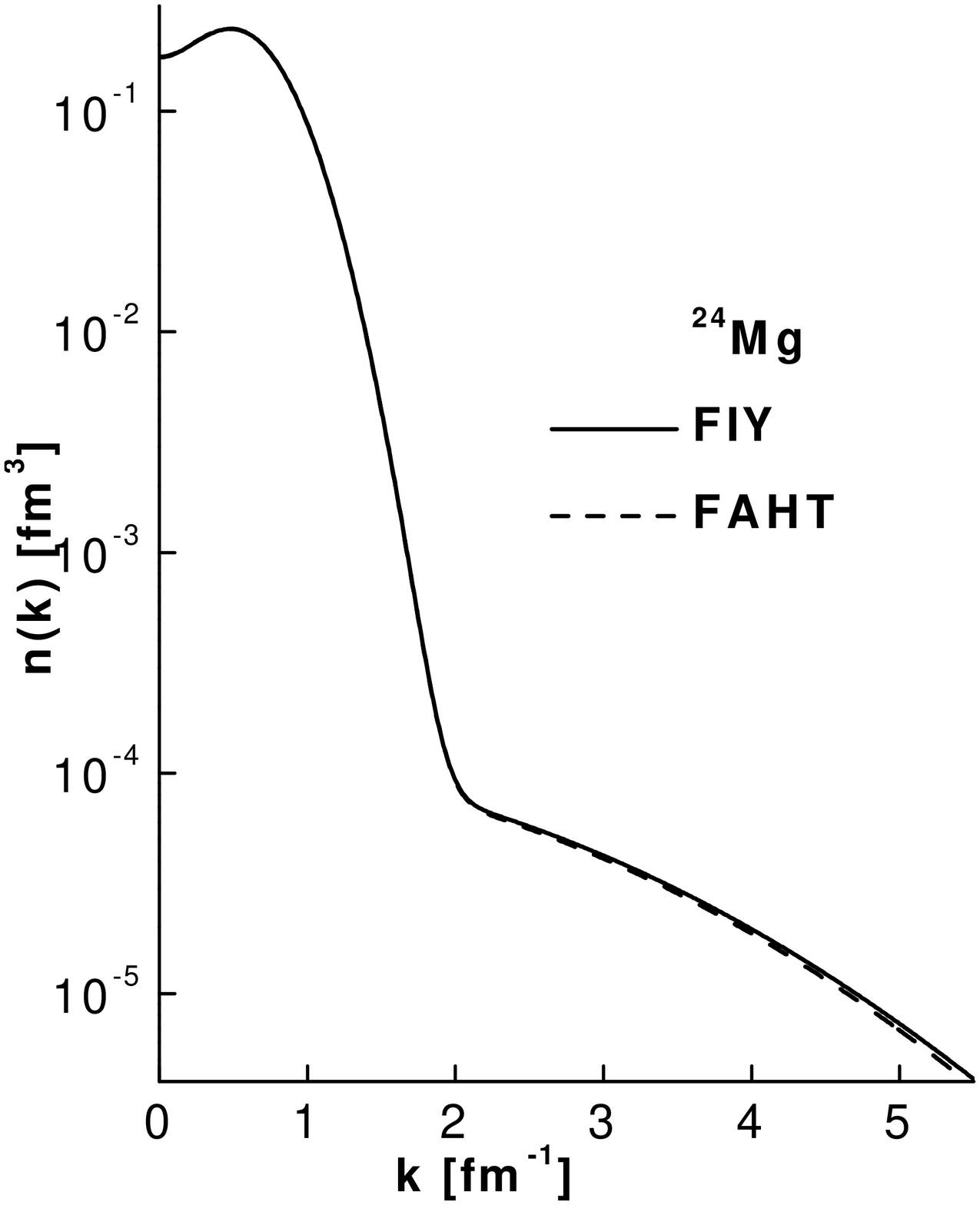,width=5cm} }  \\
&\\
&\\
&\\
{\epsfig{figure=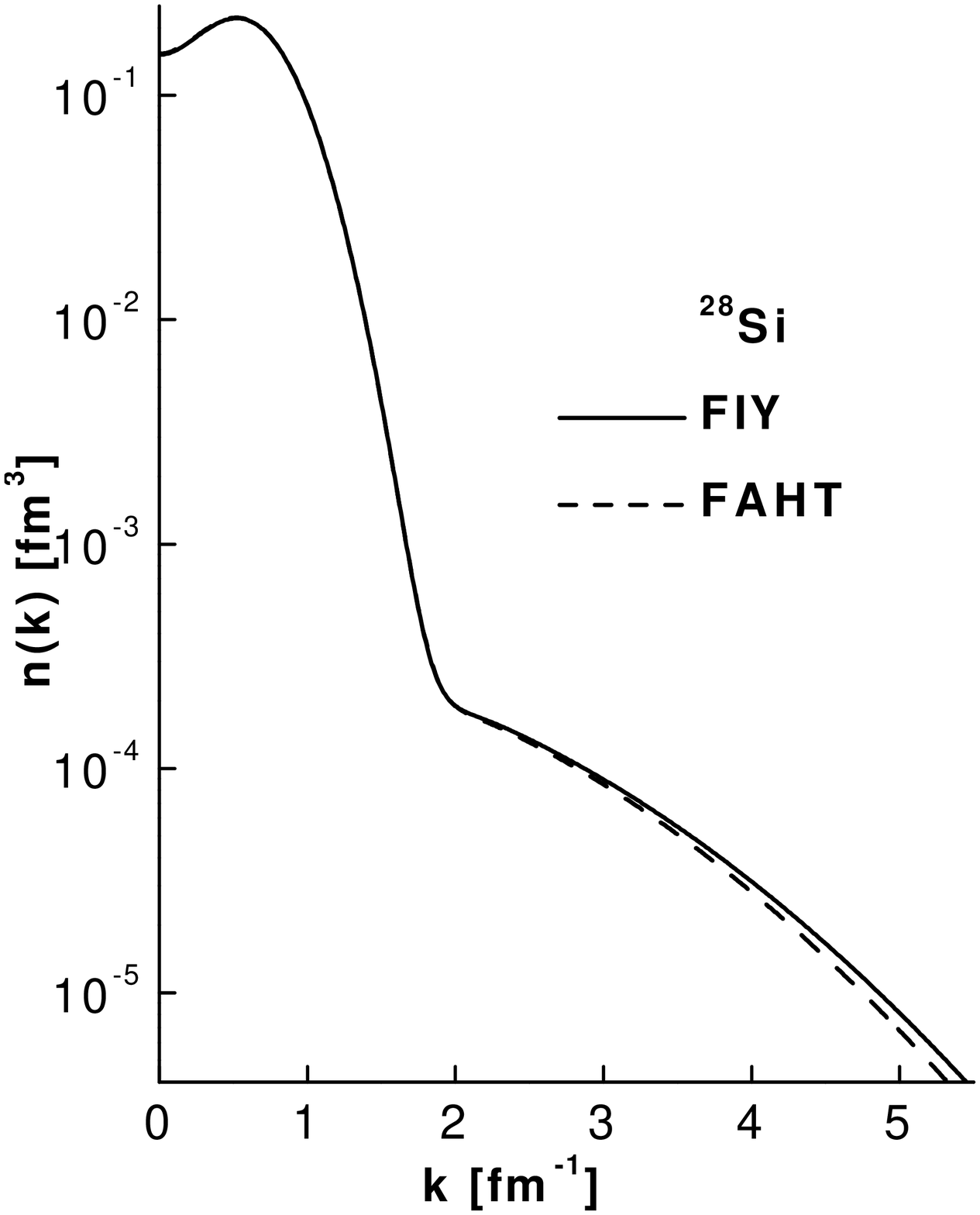,width=5cm} } &
{\epsfig{figure=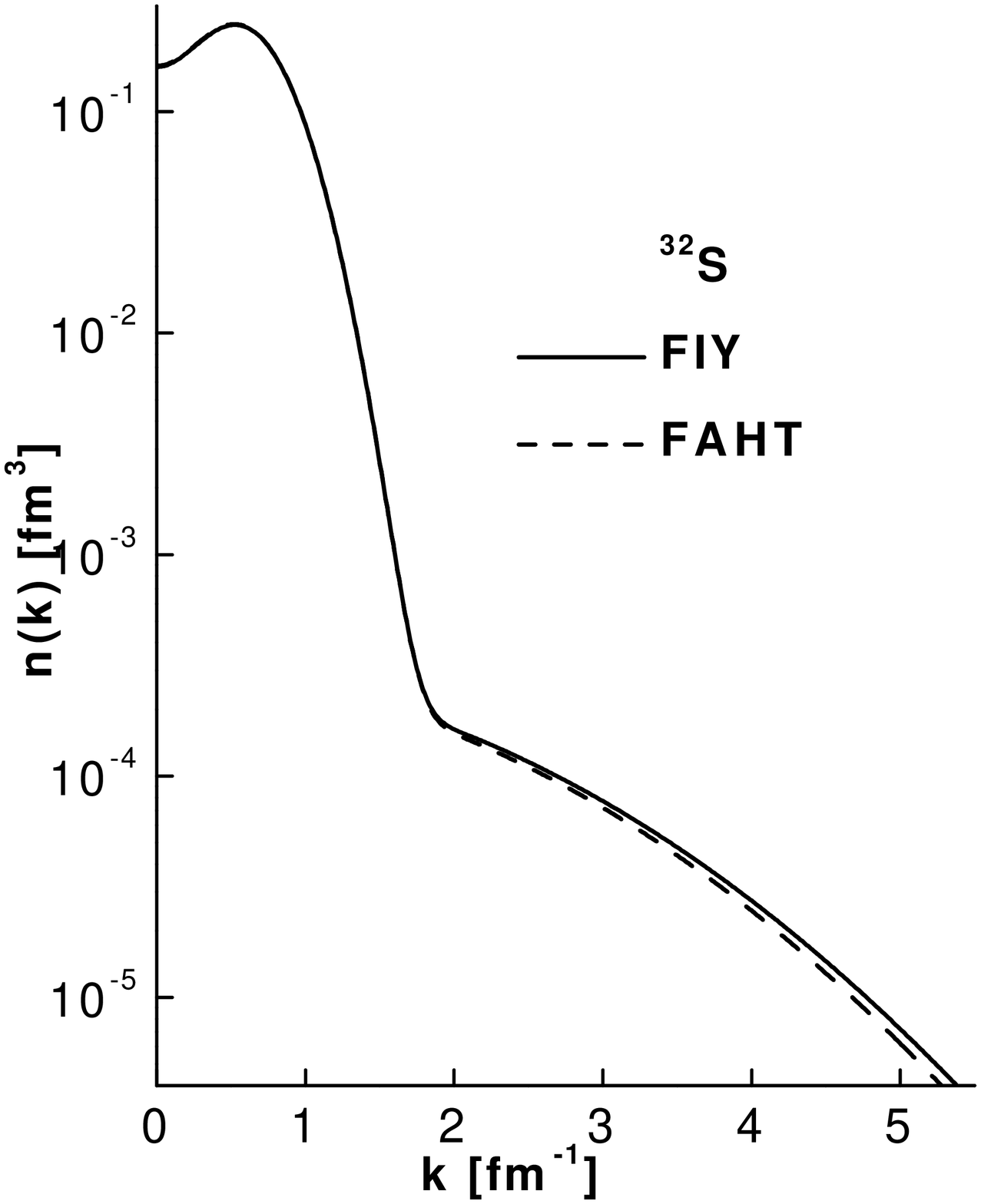,width=5cm} }  \\
\end{tabular}
\end{center}
\caption{The momentum distribution of the open shell nuclei
in the case of the FIY expansion (solid line) and the FAHT expansion
(dash line). The normalization is as in Fig. 2.}
\end{figure}

\newpage
\begin{table}
\caption{The values of the parameters $b$ (in fm) and $\beta$ (in fm$^{-2}$),
the $\chi^2$, the RMS charge radii $\langle r_{ch}^{2}\rangle^{1/2}$
(in fm), of the mean kinetic energy per nucleon $\langle T \rangle$ (in MeV)
and the nuclear information entropy in position- ($S_r$) and momentum-space
($S_k$) and the sum of them $S$ for various
$s$-$p$ and $s$-$d$ shell nuclei.
The various cases have been ordered according to increasing values
of $\chi^2$. For the various cases see text. }
\begin{center}
\begin{tabular}{l l c c r c c c c c}
%\hline
 & & & & & & & & &   \\
Nucleus &Case& $b$ & $\beta$  &
$\chi^2$ & $\langle r_{ch}^{2}\rangle^{1/2}$  &
$\langle {\bf T} \rangle$ & $S_r$ & $S_k$ & S\\
\hline
&&&&&&&&& \\
$^{4}$He&FIY   &1.1732 &2.3127  &  3.50 &1.623 &29.904 &9.978  &5.985 &15.963
\\
         &FAHT &1.1661 &1.9092  &  3.70 &1.621 &29.048 &9.943  &6.013 &15.955
\\
         &LOA  &1.1605 &1.6584  &  3.88 &1.620 &28.543 &9.917  &6.034 &15.951
\\
         &HO   &1.4320 &$\infty$& 30.94 &1.765 &15.166 &11.632 &3.014 &14.646
\\
&&&&&&&&& \\
$^{12}$C&FAHT & 1.5204 &2.4683 & 90.19 &2.427 &24.779 &31.455 &1.989  &33.444
\\
        &FIY  & 1.5190 &2.7468 & 90.87 &2.426 &25.580 &31.436 &2.142  &33.578
\\
         &HO  & 1.6251 &$\infty$&176.54 &2.490 &17.010 &32.714 &-2.2484&30.465
\\
&&&&&&&&& \\

$^{16}$O&LOA  &1.6387 &1.8825 &115.50  &2.674&23.006 &42.083 &-4.393 &37.690
\\
        &FIY  &1.6507 &2.4747 &120.19  &2.680&23.614 &42.237 &-4.557 &37.680
\\
        &FAHT &1.6554 &2.2097 &122.49  &2.684&22.518 &42.313 &-4.939 &37.374
\\
        &HO   &1.7610 &$\infty$&199.45 &2.738&15.044  &43.655 &-10.667&32.988
\\
&&&&&&&&& \\

$^{24}$Mg&FIY$^*$&1.7473 &2.4992 &140.37&3.064   &24.614 &63.532 &-14.334
&49.198
\\
        &FAHT$^*$&1.7468 &2.1833 &140.40&3.064       &23.742&63.536 &-14.603
&48.933
\\
&FIY &1.8103 &4.2275 &177.51 &3.095 &21.109&64.452 &-19.228 &45.224
\\
       &FAHT  &1.8120 &4.1322 &177.91 &3.096 &20.818 &64.483 &-19.410  &45.073
\\
       &HO    &1.8496 &$\infty$&188.01 &3.117&16.162 &65.124 &-23.429  &41.695
\\
&&&&&&&&& \\
$^{28}$Si&FAHT$^*$&1.7773 &2.1193 &103.39&3.184      &24.184 &72.901 &-20.844
&52.057
\\
          &FIY$^*$&1.7774 &2.4440 &103.47&3.184      &25.205 &72.888 &-20.438
&52.450
\\
          &FIY    &1.8236 &3.0020 &126.33 &3.216 &22.933 &73.889 &-24.115
&49.774
\\
          &FAHT  &1.8279 &2.8372 &127.84 &3.219 &22.110 &73.987 &-24.645
&49.342
\\
          &HO    &1.8941 &$\infty$&148.28&3.257 &16.099 &75.288 &-32.022
&43.266
\\
 &&&&&&&&& \\
$^{32}$S &FIY$^*$&1.8121 &2.6398 &166.11&3.282      &24.916 &82.100 &-28.343
&53.758
\\
        &FAHT$^*$&1.8131 &2.3358 &166.31& 3.283      &23.961 &82.129 &-28.827
&53.302
\\
        &FIY     &1.9368 &3.0659 &304.96  &3.443 &20.867 &86.921 &-36.707
&50.214
\\
        &FAHT   &1.9417 &2.9585 &306.46 &3.446 &20.252 &87.045 &-37.316 &49.729
\\
        &HO    &2.0016 &$\infty$&320.45 &3.483 &14.878 &88.361 &-44.881& 43.480
\\
&&&&&&&&& \\
$^{40}$Ca&FIY &1.8660 &2.1127 &160.44 &3.516 &26.617 &101.501 &-42.710 &58.791
\\
       &FAHT  &1.8685 &1.7397 &161.13 &3.517 &24.643 &101.558 &-44.172 &57.387
\\
       &LOA  &1.8164 &1.7404  &188.36 &3.397 &25.586 &97.611  &-42.121 &55.490
\\
       &HO   &1.9453 &$\infty$&229.32 &3.467 &16.437 &100.987 &-58.709&42.278
\\
%\hline
\end{tabular}
\end{center}
\end{table}

\newpage
\begin{table}
\caption{The values of the parameters $b$ (in fm), $\beta$ (in fm$^{-2}$)
and $\tilde{y}=\beta b^2$ and the values of the healing integral $w_{nl}^2$ 
for various states and for the closed shell nuclei $^4$He, $^{16}O$ and 
$^{40}$Ca and the three expansions FIY, FAHT and LOA.}
\begin{center}
\begin{tabular}{l l c c c c c c c c }
%\hline
 & & & & & & & & &  \\
Nucleus &Case& $b$ & $\beta$  & $\tilde{y}=\beta b^2$ & $w_{00}^{2}$ 
&$w_{01}^{2}$  & $w_{02}^{2}$ & $w_{10}^{2}$ & $w_{03}^{2}$  \\
\hline
 & & & & & & & & & \\
$^{4}$He &FIY & 1.1732& 2.3127&3.1832&0.01874& & & &  \\
         &FAHT& 1.1661& 1.9092&2.5961&0.02450& & & &  \\
         &LOA & 1.1605& 1.6584&2.2335&0.02971& & & &\\
& & & & & & & & & \\
$^{16}$O &FIY & 1.6507& 2.4747&6.7431 &0.00664& 0.00024&8.6$\ 10^{-6}$ 
&0.00925&   \\
         &FAHT& 1.6554& 2.2097&6.0554 &0.00773& 0.00031& 1.2$\ 10^{-5}$ 
&0.01069&  \\
         &LOA & 1.6387& 1.8825&5.0552 &0.00996& 0.00048 &2.3$\ 10^{-5}$ 
&0.01359  \\
& & & & & & & & & \\
$^{40}$Ca&FIY & 1.8660& 2.1127&7.3563 &0.00586& 0.00020&6.4$\ 10^{-6}$
&0.00821&2.1$\ 10^{-7}$  \\
         &FAHT& 1.8685& 1.7397&6.0738 &0.00770& 0.00031&1.2$\ 10^{-5}$
&0.01065&4.8$\ 10^{-7}$  \\
         &LOA & 1.8164& 1.7404&5.7421  &0.00833& 0.00035&1.5$\ 10^{-5}$
&0.01148&6.2$\ 10^{-7}$  \\
%\hline
\end{tabular}
\end{center}
\end{table}

\end{document}